# An evidence of rapid hydrogen chloride uptake on water ice in the atmosphere of Mars

Mikhail Luginin[1], Alexander Trokhimovskiy[1], Benjamin Taysum[2], Anna A. Fedorova[1], Oleg Korablev[1], Kevin S. Olsen[3], Franck Montmessin[4], Franck Lefèvre[4]

[1] Space Research Institute (IKI) RAS, Moscow, Russia
e-mail: mikhail.luginin@phystech.edu
[2] German Aerospace Center (DLR), Institute of Planetary Research, Berlin, Germany
[3] Department of Physics, University of Oxford, Oxford, UK
[4] LATMOS/CNRS, Université Paris-Saclay, Sorbonne Université, Guyancourt, France

**Abstract**

In 2020, hydrogen chloride (HCl) in the gas phase was discovered in the atmosphere of Mars with the Atmospheric Chemistry Suite (ACS) onboard the Trace Gas Orbiter (TGO) mission (Korablev et al., 2021). Its volume mixing ratio (VMR) shows a seasonal increase of up to 5 ppbv during the perihelion season, followed by a sudden drop to undetectable levels, contradicting previous estimations of the HCl lifetime of several months. In the Earth's stratosphere, heterogeneous uptake of HCl onto water ice is known to be a major sink for this species. Modelling of associated chemistry involving heterogeneous reactions indicates that $H_2O$ ice becomes the most effective sink for HCl above 20 km with the characteristic time shorter than 12 hours. In this work, we use simultaneous measurements of water ice particles and HCl abundance obtained by the ACS instrument and show particular structures in the vertical profiles, forming detached layers of gas at the ice free altitudes ("ice-holes"). We demonstrate that the heterogeneous uptake of HCl onto water ice operates on Mars and is potentially a major mechanism regulating the HCl abundance in the atmosphere of Mars.

**Main**

A search for trace atmospheric species of biological or volcanic origin is one of the primary goals of ExoMars Trace Gas Orbiter (TGO, Vago et al., 2015). To this purpose, TGO is equipped with highly sensitive infrared spectrometers, the Atmospheric Chemistry Suite (ACS; Korablev et al., 2018) and Nadir and Occultation for Mars Discovery (NOMAD; Vandaele et al., 2018) that mostly operate in solar occultation. The only new species detected so far is hydrogen chloride (HCl), a halide gas, whose signature were found in 2020 in the ACS spectra (Korablev et al., 2021) and confirmed by NOMAD (Aoki et al., 2021). Observations reveal a seasonal increase in the volume mixing ratio (VMR) up to 5 ppbv during the perihelion season, followed by a sudden drop down to undetectable values (< 0.1 ppbv) (Aoki et al., 2021; Olsen et al., 2021). The fast disappearance of HCl from the atmosphere is at variance with previous estimations of a few months HCl lifetime based on gas-phase chemistry (Aoki et al., 2021; Krasnopolsky, 2022).

The HCl abundance was mostly observed during Southern summer, a season characterised by warmer temperatures (McCleese et al., 2010), greater aerosol load (Montabone et al., 2015) and highly elevated water vapor (Maltagliati et al., 2013; Fedorova et al., 2020, 2023). Olsen et al. (2021) and Aoki et al. (2021) noticed a striking similarity between the vertical profiles of HCl and water vapor in the 5 to 25 km altitude range, suggesting that both species undergo the same atmospheric dynamics and/or that they are tied by chemistry. One possibility is that water ice could favor the heterogeneous chemical loss of HCl, which would then make the two species anticorrelated. Above the hygropause, the $H_2O$ ice clouds often form, and condensate clouds are readily detected by ACS in occultation (Fedorova et al., 2020, 2023; Luginin et al., 2020; Stcherbinine et al., 2020, 2022). The highest HCl abundances were detected during the Southern Hemisphere summer at altitudes below 20 km (Aoki et al., 2021; Olsen et al., 2021), that are almost free of water ice.

On the basis of the laboratory measurements (Kippenberger et al., 2019), both Korablev et al. (2021) and Aoki et al. (2021) speculated that HCl uptake onto water ice cloud particles could play an important role. Kippenberger et al. (2019) found that "HCl displayed extensive, continuous uptake during ice growth". Hydrogen chloride modelling in the atmosphere of Mars by Krasnopolsky (2022) also suggested a weak irreversible uptake on water ice. Taysum et al. (in prep.) have used a 1-D atmospheric photochemistry model (Taysum and Palmer, 2020) to interpret the ACS HCl observations, considering several mechanisms, including heterogeneous uptake of chlorine species onto water ice and mineral dust aerosols. This pathway has previously been used by Lefèvre et al. (2008) to model the uptake of OH and $HO_2$ radicals on water ice in the atmosphere of Mars. The modelling suggests that above 20 km, $H_2O$ ice becomes the most efficient sink for HCl.

In this work, we show observational evidence of a pronounced HCl and water ice anticorrelation above 20 km, manifested as particularly HCl-rich layers at altitudes where water ice was absent. This result supports the rapid HCl removal mechanism competing with atmospheric mixing at a diurnal rate. We used simultaneous measurements of atmospheric state and gaseous and aerosol components by ACS.

ACS is a set of three spectrometers (Korablev et al., 2018), operating in Martian orbit since 2018. Two of them were used for this study. Its Near-InfraRed (NIR) channel is an acousto-optical tunable filter echelle spectrometer working in the 0.76 to 1.6 µm range, well-suited to characterise aerosol continuum and where the $CO_2$ and $H_2O$ bands can be found. The Mid-InfraRed (MIR) channel is a cross-dispersed echelle spectrometer dedicated to solar occultation measurements in the 2.2-4.4 µm range. The echelle diffraction orders are separated along the perpendicular axis by a secondary diffraction grating. The current work focuses on diffraction orders 173–192, covering the 3.09 to 3.46 µm range. This range includes a broad water ice band encompassing all orders and strong lines of HCl in diffraction orders 173-175.

The retrieval of HCl profile abundances is done similarly to work Trokhimovskiy et al., 2021. We use the iterative Levenberg–Marquardt algorithm with Tikhonov regularisation developed to analyse SPICAM IR data on Mars Express (Fedorova et al., 2021) and ACS NIR data (Fedorova et al., 2018, 2020). We retrieve volume mixing ratios of HCl from the wavenumber range 2922.5−2927.4 $cm^{-1}$, the uncertainty on the retrieved quantities is given by the covariance matrix of the solution. The temperature and water vapor profiles used as *a priori* information are taken from available ACS NIR products (Fedorova et al., 2020, 2023).

The observed spectral range covers approximately half of the water ice absorption band centred at 3230 $cm^{–1}$, which is used to determine aerosol microphysical properties (Luginin et al., 2020; Stcherbinine et al., 2020). Simultaneous measurements of continuum absorption by ACS NIR in the spectral range 0.76-1.6 µm are included in the retrieval to increase the confidence and better distinguish between dust and water ice particles. The "onion peeling" technique (Rodgers, 2000) is used to obtain vertical profiles of the spectral dependence of aerosol extinction coefficient, which are then fitted to produce the effective radius, number density, and mass loading of water ice and dust. For details about the retrieval methodology, see Luginin et al. (2020).

We have analysed data collected by ACS during the second half of the Martian year (MY) 34 and covering two complete MY 35 and MY 36. In all, we analysed 784 occultations providing profiles of temperature, HCl (if any), water vapor, and aerosol extinction and mass loading. The three Martian years reveal a similar HCl abundance pattern (see also Olsen et al., 2021): an increase during the perihelion season, between $L_S$ = 200° and 350° when its content reaches maximum values up to ~5.5 ppbv in the Southern Hemisphere after $L_S$ = 270°. During the same $L_S$ range in the Northern Hemisphere, the HCl abundances are several times smaller. During this season, HCl-laden air masses are transported from the Southern to Northern hemisphere at high altitudes through the Hadley cell. As the northern profiles of HCl have explicit minima in the lower atmosphere (for example, see Panel a of Fig.3 in Olsen et al., 2021), it could also point to the effective gas removal in cold conditions and substantial ice load (Luginin et al., 2020).

In some occultations we observed a characteristic pattern of local anticorrelation between HCl and water ice over 5–15 km at 30–50 km altitudes. An example of this is shown in Panels A, B, and C of Figure 1. At 40-50 km, a gap in aerosol extinction at ~3 μm (water ice band) indicates an absence of water ice, while mineral dust particles can still be detected (Panels A and B). HCl is found to be maximum at altitudes where water ice is missing. This suggests rapid hydrogen chloride uptake on water ice (Panel C). Moreover, the HCl maximum is accompanied by a maximum of water vapor, that could be caused by formation of water ice above and below the "ice-hole" or a vapor release after sublimation of water ice particles. In this particluar case, the former cause is more preferable as no clear temperature maximum was observed at the altitudes of the "ice-hole" (panel B).

Out of the 137 occultations with prominent HCl abundances, 11 were found to feature clear "HCl in ice-holes" cases. These detections are localized around 30 and 45 km. Panel D of Figure 1 shows profiles of HCl VMR and aerosol extinction at 3.1 μm from those observations. In order to compare the various cases, all profiles have been normalized to [0, 1] and vertically shifted to align all HCl maxima at 45 km. The average of normalized profiles (bold lines) clearly demonstrates an anticorrelation between HCl and water ice. For nine of the 11 profiles, HCl and water ice have their maximum co-located in altitude. Over a period of three Martian years, no clear correlation with season or location was observed.

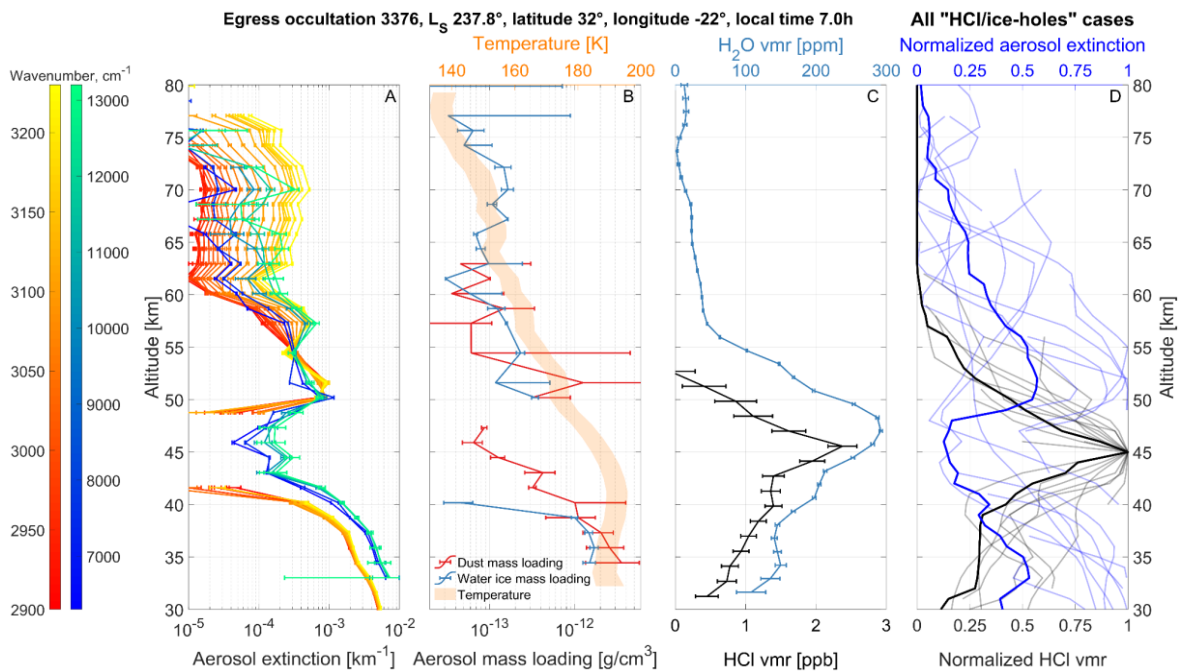

Figure 1. An illustration of the anticorrelation between the HCl abundance and water ice from a single observation (Panels A, B, and C) and from all "HCl in ice-holes" cases (Panel D). Panel A. The aerosol extinction profiles at 3.09-3.46 μm (water ice band, yellow and red) and 0.76-1.6 μm (aerosol continuum, green and blue) measured by MIR and NIR, respectively. Panel B: Retrieved aerosol mass

loading profiles of dust (red) and water ice (blue), bottom axes; temperature profile with uncertainties retrieved from NIR data (Fedorova et al., 2023), orange area and top axes. Panel C: volume mixing ratio of HCl (black lines and bottom axes) and $H_2O$ (blue lines and top axes) retrieved from MIR and NIR, respectively. Panel D: profiles of HCl VMR (black lines) and aerosol extinction at 3.1 μm (blue lines) normalized to [0, 1] and shifted to align all HCl maxima at 45 km; mean profiles are shown in bold. Error bars of the experimental data correspond to 1−σ uncertainty level.

As mentioned above, the heterogeneous uptake of HCl onto water ice is a process known to occur in the Earth's atmosphere, acting as a major sink for chlorine in the stratosphere (Huthwelker et al., 2006). Qualitatively, the reactions are known to be very efficient with uptake coefficients (probability of a HCl molecule reacting or absorbing onto the ice crystal surface) experimentally measured with values ≥ 0.1 (Burkholder et al., 2019). The quantitative values and precise mechanisms acting at the surface are still debated and highly depend on the presence of other gas-phase acid species, the ice surface roughness, atmospheric temperatures, and HCl partial pressures (Abbatt, 1997; Barone et al., 1999; Chu et al., 1993; Fernandez et al., 2005). Gas-phase chemistry on Mars predicts HCl lifetimes of a few months (Aoki et al., 2021; Krasnopolsky, 2022). Hydroxyl (OH) radical reactions dominate below the hygropause, photolysis is most efficient between the hygropause and ~ 40 km, and hydrogen (H) atoms become the strongest sink in the upper (> 40 km) atmosphere. Taysum et al. (in prep.) use a temperature-dependent uptake coefficient from Hynes et al. (2001) scaled by a non-dissociative Langmuir coefficient (McNeill et al., 2007) to account for saturation of uptake sites on the ice surface and find that the presence of water ice can lower this lifetime to timescales less than 1 hour. The model finds that heterogeneous loss on ice can be over two orders of magnitude greater than gas-phase processes below the hygropause. For the observation in Figure 1, calculations for altitudes 35-40 km predict the HCl lifetime to range from 1.5 to12 hours. The presence of the "ice-holes" within the clouds thus enables the build-up of HCl within these ice-free regions, and drives the development of a distinct layered structure in the HCl profiles by removing it above and below.

The study presented here is a first, yet robust, exploration of the active connection between hydrogen chloride and water ice in the martian atmosphere. We propose that the uptake of HCl on water ice efficiently and rapidly removes hydrogen chloride from the atmosphere and probably restricts it from reaching higher altitudes. The fact that HCl, when detected, is found to be maximum in the "holes" of the ice profiles reveals the preservation of HCl at these altitudes and offers observational evidence of chemical interactions between HCl and water ice on a diurnal scale.

**Acknowledgments**

This work has been funded by grant #23-12-00207 of the Russian Science Foundation. KSO acknowledges funding from the UK Space Agency (ST/T002069/1, ST/Y000196/1). FM and FL acknowledge funding from CNES.